\date{May 20, 1999}

\documentstyle[12pt]{article}

\catcode`\@=11
\newcommand{\ccite}[1]
{\@ifundefined{b@#1}{\bf ?}{\@nameuse{b@#1}}}

\begin{document}
\baselineskip = 20pt

\title{The effect of a topological gauge field on 
Bose-Einstein condensation}

\author{P. F. Borges,\thanks{e-mail: pborges@if.ufrj.br;} 
H. Boschi-Filho\thanks{e-mail: boschi@if.ufrj.br;} 
and C. Farina\thanks{e-mail: farina@if.ufrj.br.}
\\ 
\small \it
Instituto de F\'\i sica, Universidade Federal do Rio de Janeiro \\ 
\small \it Cidade Universit\'aria, Ilha do Fund\~ao, Caixa
Postal 68528 \\ 
\small \it 21945-970 Rio de Janeiro, BRAZIL}


\maketitle

\begin{abstract} 
We show that Bose-Einstein condensation 
of charged  scalar fields 
interacting with a topological gauge field at finite temperature 
is inhibited except for special values of the topological field. 
We also show that fermions interacting
with this topological gauge field can condense for some values
of the gauge field.
\end{abstract}

\vfill


\vskip 1cm

\pagebreak

\section{Introduction}

Bose-Einstein condensation (BEC) has attracted much attention
since its proposal by Einstein \cite{Einstein}, after the
fundamental work of Bose \cite{Bose} on the statistics of integer
spin particles. Its first realization was given by London in a
model for the Helium superfluidity \cite{London}. More recently,
the interest for this phenomenon has increased because of the
experimental results on atomic traps and the direct observation
of Bose-Einstein condensation (BEC) on gases \cite{exp}
signalizing for a large number of applications. 
In view of these aspects, it has been a subject closely 
related to condensed matter physics. 
However, this situation has changed
over the past two decades since the work of Haber and Weldon
\cite{HW} and Kapusta \cite{Kapusta} where BEC was related to
spontaneous symmetry breaking in relativistic quantum field
theory at finite temperature and density. 
Long ago, it was shown for massive bosons in $N$ 
space dimensions and in the absence of
external fields that BEC can only 
occur\footnote{For massless bosons this
condition reduces to $N \ge 2$, although in this case the
critical temperature is infinite.} if $N \ge 3$. 
This result was first shown
in a nonrelativistic context \cite{Huang} and later generalized to 
the relativistic case [\ccite{HW},\ccite{Kapusta}]. Recently, the
extension of these results to curved space-times and the
inclusion of non-uniform magnetic fields in the relativistic
context were also done \cite{Toms,KirstenToms}.

The topology of finite temperature field theory is nontrivial due
to the compactification of the time coordinate (after a Euclidean
rotation) to a circle of length $\beta=T^{-1}$. This topology 
and the periodicity of bosonic fields at finite temperature
imply that even being constant and uniform the 
time component of the gauge field $A_0$ 
cannot be gauged away from the theory, but just be
reduced to a value in the interval $(0,2\pi/\beta)$, 
except in the case where $A_0=2\pi n/\beta$ ($n$~integer) 
for which it can be reduced to zero. 
The relevance of $A_0$ is manifold. 
For instance, it became clear after the work of Polyakov 
\cite{Polyakov78} and Susskind \cite{Susskind79} 
that the problem of confinement in gauge theories 
at finite temperatures, 
which in general can be inferred from the expected 
value of the Wilson line $<\exp i\int dx_\mu A^\mu >$, 
can be analyzed simply looking at 
$L=<\exp \int_0^\beta d\tau A_0 >$. 
If $L=0$ then the charge will be confined and 
in the other case it is deconfined 
(see also [\ccite{Gava80}-\ccite{GrossPY80}]). 
In general, $A_0$ is a function of space-time coordinates 
$(\tau,\vec x)$ but the most simple case where an 
effective potential $V(A_0)$ can be calculated is the 
case of a constant and uniform $A_0$ as has been discussed in Refs. 
[\ccite{GrossPY80}-\ccite{Actor83}] (of course for slowly varying 
$A_0$ this can be viewed as a good approximation). 
In particular, scalar and spinor 
electrodynamics at finite temperature in $N+1$ space-time 
dimensions with $A_0\approx$ constant 
have been studied by Actor \cite{Actor83} 
where the role of $A_0$ as a ``thermal interaction'' 
has been stressed. He also considered abelian gauge 
theories in the presence of constant $A_0$ 
at finite temperature and density, 
which means a nonzero chemical potential. 
Recently, in 2+1 dimensional physics it was shown that 
the presence of an $A_0$ is essential to keep gauge 
invariance of the Chern-Simons term 
at finite temperature \cite{Dunne97}.
However, the influence of such a gauge field 
(constant and uniform $A_0$) in BEC of a gas of 
relativistic charged bosons has not been investigated. 

Here, we present a relativistic quantum field
theory approach to BEC of charged scalar fields interacting
with a topological gauge field $A_\mu$ in $N+1$ space-time dimensions at
finite temperature and density. For calculational purposes we take
$A_i=0$ and $A_0\approx$ constant following the lines of Refs. 
[\ccite{GrossPY80}-\ccite{Actor83}]. 
The case $A_i\not= 0$ will be discussed in a separate paper. 
We show that BEC is inhibited except for special values of $A_0$. 
Moreover, we also show that such a gauge field interacting with
a gas of relativistic charged fermions can induce condensation analogous 
to BEC for some other values of the gauge field. 
These results are also valid in nonrelativistic context
which one can obtain from our relativistic results as a particular limit.

\section{Bose-Einstein Condensation}

The interaction of charged scalar fields with $A_0$ is equivalent
to modifying the usual periodic (bosonic) boundary conditions 
for the charged scalar fields at
finite temperature to generalized boundary conditions
\cite{Actor83,Sachs,BBF} or to the introduction of an
imaginary part of the chemical potential (this will be made
explicity bellow). An imaginary chemical potential has been
considered recently in the literature to discuss the critical
exponents of the Gross-Neveu model in 2+1 dimensions at fixed
fermion number \cite{Nogueira} and the potential for the 
phase of the Wilson line for an $SU(N)$ gauge theory 
at nonzero quark density \cite{Pisarski99}.
In these two papers it was suggested that an imaginary part 
of the chemical potential allow fermions to condense. 
In the next section, we are going to show this result explicitly
for a gas of relativistic charged fermions.

Let us start writing the partition function for the charged 
scalar fields $\Phi$ and $\Phi^\ast$ with mass $M$ 
at finite temperature and density interacting with the 
topological gauge field $A_\mu=(A_0,\vec 0)$ 
in $N+1$ space-time dimensions as 
(see \cite{Kapusta,Actor83} for details)
\begin{equation}\label{Zboson}
{\cal Z}=\int_{periodic} D\Phi D\Phi^\ast
\exp\left\{-\int_0^\beta d\tau \int d^Nx\  
\Phi^\ast[(\partial_0 +A_0 +i\mu)^2 -\partial_i^ 2 + M^2] \Phi
\right\}
\end{equation}

\noindent where $\mu$ is the chemical potential and 
Latin letters indicate spatial coordinates
\hbox{$(i=1,...,N)$}. As it is well known, one can
express this partition function as a determinant namely,
\begin{equation}\label{bosondet}
{\cal Z}_{Bosons}=\left\{\left.
\det[-(\partial_0 + A_0 + i\mu)^2-(\partial_i)^2 + M^2]
\right|_{P}\right\}^{-1}\;.
\end{equation}

\noindent The label {\it P} means that the
eigenvalues of this operator are subjected to periodic boundary
conditions and hence they are given by
\begin{equation}\label{eigenvalues} 
\lambda_{nk}=(\omega_n + A_0 +
i\mu)^2+\vec k^2+M^2\;, \end{equation}

\noindent where $\omega_n=2n\pi/\beta$, with $n\in {\sf Z}\!\!
{\sf Z}$, are the Matsubara frequencies for bosonic fields and
$\vec k\, \in \, {{\sf I}\!\!\; {\sf R}}^N$. 
The above determinant can be calculated by different methods as for
example the zeta function [\ccite{Hawking77}-\ccite{Weldon86}] using 
\begin{equation}
\det {\cal O}=\exp
\left[(-\partial/\partial s) \zeta (s, {\cal O})
\right]_{s=0}\;,
\end{equation}
where the generalized zeta function is given by
\begin{equation}
\zeta (s, {\cal O})
=\sum_n \lambda_{n}^{-s}\;,
\end{equation}

\noindent and $\lambda_{n}$ are the eigenvalues of the 
operator ${\cal O}$. An analytical extension of 
$\zeta(s,{\cal O})$ for the whole s-complex plane is assumed.
\noindent Then, it can be shown that
the free energy $\Omega=-(1/\beta)\ln{\cal Z}$ corresponding to
the partition function (\ref{Zboson}) can be written
as\footnote{Note that in this formula, only the real part of the
free energy $\Omega(\beta,\mu,A_0)$ is expressed, according to the
prescription of introducing the chemical potential as an
imaginary time-component gauge potential.} \cite{Actor83}
\begin{eqnarray} 
\Omega(\beta,\mu,A_0)= - 4 V
\sum_{n=1}^{+\infty}\cos(n\beta A_0)\cosh(n\beta\mu) 
\left({M\over 2\pi n\beta}\right)^{N+1\over 2}
K_{{1 \over 2} (N+1)}(n\beta M)\;, 
\label{Omega}
\end{eqnarray}

\noindent where $V$ is the volume in $N$ space dimensions and 
$K_\nu(x)$ is the modified Bessel function of the second kind. 
Further, the charge density is given by:
\begin{equation}\label{rho}
\rho\equiv {1\over \beta V}{\partial\over\partial\mu} \ln {\cal Z} 
= -{1\over V}\left({\partial \Omega\over\partial\mu}\right)_{V,T}.
\end{equation}

\noindent 
We can use the above  free energy to calculate the critical 
temperatures, densities and dimensions for BEC. 
Let us illustrate this with the simplest cases where we take 
$A_0=0$ and the limits of ultrarelativistic or nonrelativistic 
regimes. The general case $A_0\not=0$ will be discussed bellow.
Let us first consider the ultrarelativistic case. In this
situation where the energies involved are much higher than the
mass scale $M$ we should take the limit $\beta M<<1$, which is
easily recognized as a high temperature limit. The condensation
condition $\mu\rightarrow M$ also implies $\beta \mu<<1$, so 
that the charge density reads (see the Appendix)
\begin{equation}
\label{highrho} 
\rho = {2\mu \over
\pi^{(N+1)/2}}\Gamma({N+1\over 2}) \zeta(N-1) T^{N-1} , 
\end{equation}

\noindent where $\zeta(s)=\sum_{n=1}^\infty n^{-s}$ is the usual
Riemann zeta function. Using the fact that 
the condensate is reached when $\mu=M$, in $N=3$
space dimensions, we find the critical temperature:
\begin{equation}
T_c=\left( {3\rho\over M}\right)^{1/2},
\end{equation}

\noindent which 
coincides with the ultrarelativistic BEC results
known in the literature [\ccite{HW}-\ccite{KirstenToms}]. 
Now, let us discuss briefly the situation for the nonrelativistic 
limit which
here means to take the low temperature limit, $\beta M >>1$. 
\noindent At the condensate $\mu=M$, 
the charge density is given by 
\begin{equation}\label{lowrho}
\rho = \zeta({N\over 2})\left({T_c M\over 2\pi}\right)^{N\over 2}, 
\end{equation}

\noindent which for $N=3$ 
space dimensions leads to the critical temperature 
\begin{equation}
T_c={2\pi\over M} \left({\rho\over \zeta(3/2)}\right)^{2/3},
\end{equation}

\noindent which is in agreement with the well known 
nonrelativistic BEC results [\ccite{Huang}-\ccite{KirstenToms}].
Looking at the Eqs. (\ref{highrho}) and (\ref{lowrho}) for the
charge density in $N$ space dimensions for the ultrarelativistic
and nonrelativistic cases one can see that the condensate is not
defined in two space dimensions for both cases, since the Riemann
zeta function $\zeta (s)$ has a pole at $s=1$, so that condensation 
occurs for $N>2$.


Let us now discuss the general case $A_0\not=0$.
To see explicitly when the condensation occurs, let us rewrite
the free energy (\ref{Omega}) as 
\begin{eqnarray}
\Omega(\beta,\mu,A_0)
={TV \over (2\pi)^N} \int d^N k \;\; \Re
\left\{ \ln\left[1-e^{i\beta A_0} e^{-\beta(\omega-\mu)}\right] +
(\mu\rightarrow -\mu) \right\} \qquad\qquad\nonumber \\
 ={TV \over 2 (2\pi)^N} \int d^N k 
\left\{ \ln\left[1+e^{-2\beta(\omega-\mu)}
-2\cos(\beta A_0)\, e^{-\beta(\omega-\mu)}\right]
+ (\mu\rightarrow -\mu)
\right\}\;, \label{Omegaprime}
\end{eqnarray}

\noindent where $\omega=\sqrt{\vec k^2 + M^2}$. 
The condensation condition is given by the divergence 
in the free energy or equivalently in the charge density. 
As the condensation occurs near the zero momenta
($\vec k=\vec 0$) state, this implies that $\omega\rightarrow M$
and the condensation condition is given by 
\begin{equation} \label{condition1}
1+e^{-2\beta(M-\mu)}-2\cos(\beta A_0)\;
e^{-\beta(M-\mu)}=0\;, 
\end{equation}

\noindent or simply 
\begin{equation}\label{condition2}
\cosh\beta(\mu -M)=\cos\beta A_0\;,
\end{equation}

\noindent which can only be satisfied for finite temperatures 
if $\mu=M$ and
\begin{equation}\label{BECA0}
A_0={2n\pi\over\beta} \qquad\qquad (n\in {\sf Z}\!\! {\sf Z})
\end{equation}

\noindent simultaneously. 
Note that these values for $A_0$ are precisely the ones that can
be gauged to zero. Hence, we can say that BEC 
is completely suppressed
for $A_0\not=0$. The above condition on $A_0$
coincides with the minima of the free energy Eq. (\ref{Omega})
indicating that the values where condensation occurs are minima
for interaction energy of the charged   scalar fields with
$A_0$. This condition on the topological field $A_0$ also
implies that the trivial ($n=0$) topological sector  is equivalent
to the zero temperature limit $(\beta\to\infty)$, as expected.

As a final remark on BEC of charged bosonic particles, let us mention that
the divergence condition given by Eqs. (\ref{condition1}) 
and (\ref{condition2}) comes from the branch cut of the free energy 
(\ref{Omegaprime}) defined by $\mu^2_{\rm cut}\le M^2$, 
when $A_0$ satisfies Eq. (\ref{BECA0}). 
The physical region for the chemical potential
is then $\mu^2> M^2$ and the limit $\mu\to M$ defines the critical point 
for which BEC happens.

\section{Fermions}

Let us now show that the above argument for bosonic scalar fields
applied to fermionic fields implies that for special values of the
topological gauge field $A_0$ fermions can condense. 
The partition function in this case can be written as
\begin{equation}\label{Zfermion}
{\cal Z}^F=\int_{antiper.} D\bar\Psi D\Psi
\exp\left\{\int_0^\beta d\tau \int d^Nx\  
\bar\Psi [-i\gamma_0(\partial_0 +A_0 +i\mu) 
-i\gamma_i\partial_i - M] \Psi
\right\}.
\end{equation}

\noindent As for the
case of bosons, one can write the fermionic partition 
function as a determinant \cite{Kapusta,Actor83} 
\begin{eqnarray}\label{fermiondet}
{\cal Z}_{Fermions} 
& = & \left.{\det}_D(i\slash\!\!\!\!D -M)\right|_{A}
\nonumber\\ 
& = & \left[\left.\det(-D^2+M^2)\right|_{A}\right]^{+1}\;,
\end{eqnarray}

\noindent where $\det_D$ means the calculation over
Dirac indices and the subscript $A$ means that the eigenvalues of
each operator are computed with antiperiodic boundary conditions.
Hence, for the operator $-D^2+M^2$, the eigenvalues are given by 
(\ref{eigenvalues}), but now the Matsubara frequencies are 
$\omega_n=(2\pi/\beta)(n+1/2)$ with $n\in {\sf Z}\!\! {\sf Z}$, 
so that
\begin{equation}\label{eigenvalues2}
\lambda^F_{n k} = 
\left[{(2n+1)\pi \over \beta} +i\mu + A_0 \right]^2 
+\vec k^2+M^2\;,
\end {equation}

\noindent and the corresponding free energy is given by 
\cite{Actor83}
\begin{eqnarray} 
\Omega_F(\beta,\mu,A_0) = 4 V
\sum_{n=1}^{+\infty}\cos[n(\pi+\beta A_0)]\cosh(n\beta\mu) 
\left({ M\over 2\pi n\beta}\right)^{N+1\over 2}
K_{{1 \over 2} (N+1)}(n\beta M)\;,
\label{Omegaf}
\end{eqnarray}

\noindent which can also be written as
\begin{eqnarray}
\Omega_F(\beta,\mu,A_0)
=-{TV\over 2(2\pi)^N}\int{d^N k}
\Big\{\ln \left[1+e^{-2\beta(\omega-\mu)}
-2\cos(\pi+\beta A_0)e^{-\beta(\omega-\mu)}\right]
\nonumber\\
+(\mu\rightarrow-\mu)\Big\}.
\label{Omegaprimef}
\end{eqnarray}

\noindent Noting the similarity of Eqs. (\ref{Omegaf}), 
(\ref{Omegaprimef}) for the fermionic free energy with Eqs. 
 (\ref{Omega}), (\ref{Omegaprime}) for the bosonic case  
we see that the previous analysis for the bosonic 
 scalar field follows immediately with minor changes. 
Here, the condensation condition is modified to 
\begin{equation}
\cosh\beta(\mu -M)=\cos(\pi + \beta A_0)\;,
\end{equation}

\noindent so that we see that fermions can condense when $\mu=M$
analogous to bosons, but the topological gauge field should assume 
the values 
\begin{equation}
{A_0}^\prime={(2n+1)\pi\over\beta}, \qquad\qquad 
(n\in {\sf Z}\!\! {\sf Z}).
\end{equation}

\noindent Note that these values for $A_0^\prime$ 
cannot be put to zero
by a gauge transformation $\Delta A_0=2n\pi/\beta$, 
because of (anti)periodic boundary
conditions of (fermion) gauge field.
In this case in the trivial ($n=0$) topological sector 
 $A_0^\prime$ is non vanishing which value $\pi/\beta$ we 
interpret as the exact amount to transmutate the fermion 
into a boson.
The values for $A_0^\prime$ correspond to the {\sl
maxima} of the free energy Eq. (\ref{Omegaf}), which is exactly
minus the bosonic free energy Eq. (\ref{Omega}). This relation
between the free energies of the bosonic and fermionic cases was 
anticipated by Actor \cite{Actor83}, although without discussing BEC. 
The remaining discussion on condensation is unaltered unless for the
substitution $A_0\to A_0^\prime$ and the critical temperatures
and dimensions are unchanged.

\section{Conclusion}

We described here a kind of finite temperature bosonization (or
fermionization) by including the topological gauge field $A_0$ in
the partition function. We have shown that no condensation occurs
for charged scalar fields coupled to a constant and uniform 
gauge field $(A_0,\vec 0)$ 
with $ A_0\beta\not=2n\pi$. However, a charged 
fermionic field interacting with a topological field
$A_0=(2n+1)\pi / \beta$ will also condense, so this should
correspond to a bosonization process. Reversely, adding such a
field to a boson we fermionize it once they will exclude rather
than condense. This picture is valid only at finite
temperatures since at zero temperature $(\beta\to\infty)$ 
the topological field
$A_0$ which allows BEC vanishes. 

This fermionization/bosonization process at finite temperature 
can also be seen from the 
$A_0$-dependent distribution function corresponding to 
the bosonic free energy (\ref{Omega}) which we write as:
\begin{eqnarray}
f_{A_0}(\beta,\mu) 
= \Re \left\{ {1\over e^{i\beta A_0}e^{\beta(\omega-\mu)}-1}
	\right\}.
\end{eqnarray}

\noindent When $\beta A_0=2n\pi$ we obtain the Bose-Einstein
distribution function 
\begin{eqnarray}
f_0(\beta,\mu) 
= {1\over e^{\beta(\omega-\mu)}-1}
\end{eqnarray}

\noindent and when $\beta A_0=(2n+1)\pi$ we get
\begin{eqnarray}
f_{\pi}(\beta,\mu) 
= - {1\over e^{\beta(\omega-\mu)}+1} ,
\end{eqnarray}

\noindent which is minus the Fermi-Dirac distribution function,
so in some sense the boson interacting with $A_0=(2n+1)\pi/\beta$
corresponds to an antifermion. Analogously, we can analyze the
case where fermions interact with the topological field $A_0$
giving rise to the distribution function:
\begin{eqnarray}
f^F_{A_0}(\beta,\mu) 
= \Re \left\{ {1\over e^{i\beta A_0}e^{\beta(\omega-\mu)}+1}
	\right\}.
\end{eqnarray}

\noindent In this case, if $\beta A_0=2n\pi$ we find the
Fermi-Dirac distribution function, $-f_\pi(\beta,\mu)$, and if
$\beta A_0=(2n+1)\pi$ the distribution function is minus the
Bose-Einstein one, $-f_0(\beta,\mu)$, and the fermion is
transmutated into an antiboson. For noninteger values of $\beta
A_0/\pi$ these distribution functions may describe interpolating
statistics between Fermi-Dirac and Bose-Einstein ones as
discussed in \cite{BBF}.

The results of transmutation of fermions into
bosons found here for a gas of relativistic charged fermions 
are similar to those found in \cite{Nogueira} within the 
2+1 dimensional Gross-Neveu model and in \cite{Pisarski99} 
for an $SU(N)$ gauge theory at nonzero quark density where 
both considered an imaginary chemical potential. 
From our analysis, we also expect that a similar study of
bosonic models at finite temperature
with an imaginary chemical potential should
present transmutation of bosons into fermions.

\bigskip
\bigskip

\noindent {\bf Acknowledgments.} 
The authors acknowledge interesting discussions with M. B. Silva Neto 
and for calling our attention to Refs. \cite{Nogueira} and 
\cite{Pisarski99}. H.B.-F. thanks the hospitality of MIT-CTP where part
of this work was done. H.B.-F. and C.F. were partially supported by CNPq
(Brazilian agency).



\section{Appendix: High and low temperature limits}

Here for self-completeness we show some details on the high and
low temperature limits relevant for BEC when the free energy is 
expressed in terms of Bessel functions, as in Eq. (\ref{Omega}). 
In the ultrarelativistic
regime $\beta M<<1$ we can use the following property of the
Bessel functions
\begin{equation}
\lim_{x\rightarrow 0} K_\nu (x) = \Gamma(\nu) 2^{\nu-1} x^{-\nu}
\;;\;\;\;\;\;\;\;\;\;\;\;\;\;\;\;(\nu>0)
\end{equation}

\noindent so that the free energy (\ref{Omega}) 
reduces to ($\beta A_0=2n\pi$)
\begin{equation}
\Omega(\beta,\mu)=-{2V\over (\beta \sqrt\pi)^{N+1}}
\Gamma({N+1\over 2})\sum_{n=1}^{+\infty}\cosh(n\beta\mu)
({1 \over n})^{N+1}\;.
\label{highOmega}
\end{equation}

\noindent From the definition of the charge density 
Eq. (\ref{rho}) we find that  at 
high temperature it is given by
\begin{equation}
\rho(\beta,\mu)={2 \over (\sqrt\pi)^{N+1}} 
\left({1\over \beta}\right)^N \Gamma({N+1\over 2})
\sum_{n=1}^{+\infty}\sinh(n\beta\mu)
({1 \over n})^{N}\;.
\end{equation}

\noindent From this charge density one can find the 
Eq. (\ref{highrho}).

For the nonrelativistic limit we can use 
the asymptotic expansion for the
Bessel function
\begin{equation}\label{asymptotic}
K_\nu (x) \simeq \sqrt{\pi\over 2x} e^{-x},
\end{equation}

\noindent valid when $|x|>>1$ and $-\pi/2<\arg x<\pi/2$. 
Taking the limit $\beta M >>1$ in the free
energy(\ref{Omega}), we have in this case ($\beta A_0=2n\pi$): 
\begin{equation}
\Omega(\beta,\mu)=-{2V\over\beta} \left({M \over
2\pi\beta}\right)^{N\over 2} 
\sum_{n=1}^{+\infty}\cosh(n\beta\mu) 
({1 \over n})^{1+{N\over 2}}e^{-n\beta M}\;. 
\end{equation}

\noindent In this regime the charge density reads
\begin{equation}
\rho(\beta,\mu)= 2 \left({M \over 2\pi\beta}\right)^{N\over 2}
\sum_{n=1}^{+\infty}\sinh(n\beta\mu) ({1 \over
n})^{{N\over 2}}e^{-n\beta M}\;. 
\end{equation}

\noindent Exactly as in the ultrarelativistic case, here the
critical point is reached when $\mu\rightarrow M$. So, in the
nonrelativistic limit, the condensation condition implies that
$\beta\mu >>1$ and then the charge density reduces to 
\begin{equation}
\rho(\beta,\mu)= 
\left({M \over 2\pi\beta}\right)^{N\over 2}
\sum_{n=1}^{+\infty} ({1 \over n})^{{N\over 2}} 
e^{n\beta(\mu- M)},
\end{equation}

\noindent from which we find Eq. (\ref{lowrho}).


\newpage

\end{document}